# Eliminating excess phase accumulation in a continuous perturbed heterogeneous planar PhC

SHAHRAM MORADI[1, 2]

1Electrical and Computer Engineering, University of Victoria, Victoria, Canada
2Micro and Nano Technology Department, Middle East Technical University, Ankara, Turkey
Email: photonicsandoptics@gmail.com

In this paper, we propose an asymmetric distribution of a hexagonal lattice for achieving near-zero group velocity with negative group delay. This study reports the effect of the continuous geometric perturbation on the photonic band diagram and consequently its impact on phase velocity, group velocity, and effective refractive index. We provide a promising method for modifying the photonic band diagram for obtaining an exotic dispersion diagram. With this obtaining a broadband spanning from E to L band that envelop of light pulse traverse with almost zero velocity is promising enough to apply it in a wide variety of light-based devices.

**Introduction**

Group velocity $(v_g)$ is the phase response of a medium that is computed through the derivative of phase velocity $v_g$ with respect to a certain range of frequencies. A light pulse containing series of sine waves (a particular range of band frequencies) that each experiences different phase response can be steered via dexterous engineering matter. Dispersion diagram as a result of this diversity in phase response in a man-made matter provides a useful information that explains how a light pulse envelope interacts with matter during propagation in a transmission path. Three types of dispersing occurs including non-dispersive ($v_g = v_p$), normal dispersive ($v_g < v_p$) and anomalous dispersive ($v_g > v_p$) in various types of optical components including artificial and/or normal mediums. In terms of application, shortening or reducing transmission delays have capability of playing a critical role in ultra-high optical switching elements[1] and today's complex modulation techniques in modern RF wireless communicating platforms[1]. In addition, all-optical regeneration of a signal[2] is desired to avoid polarization mode dispersion[3], [4], eliminate influences of noise and fade limitations associated with wavelength-division multiplexed (WDM) transmission[5]. In fact, conducting light pulse through a chromatic dispersive material with high accuracy, low loss and no distortion in fiber optics and/or any on-chip elements is a hard task. However, advance shift in phase compensate positive group delay of the transmission line and also assist to improve efficiency of amplifiers [6], [7].

Negative Group Delay (NGD)[8] as a unique phenomenon gives rise to advance in phase which does not occur in majority of transmission lines of EM waves. Considering a traversing electromagnetic wave through a dispersive matter, an envelope of magnitude experience an advance shift rather than delay via negative group delay. This promising effect have been studied in both electronic circuitry scope[9], [10] and even in high frequency structures[11], [12]. By considering $e^{iK(\omega)z}$ as a transfer function of normalized plane wave transmitting in the **z** axis, its complex wave number is $k(\omega) = \alpha(\omega) + i\beta(\omega)$. The corresponding phase and group delays at any particular point in **z** axis are:

$$T_p = \frac{z}{v_p} = \frac{\alpha(\omega)}{\omega}z, \ \ T_g = \frac{z}{v_g} = \frac{d}{d\omega}(\alpha(\omega)).z$$

, where $v_p, v_g$ are the phase and group velocities respectively. Furthermore, according to the Sommerfield, the front end of the wave packet with the velocity of $v_f$ equals to light speed ($c$) in vacuum since:

$$T_f = \frac{z}{v_f} = \lim_{\omega \to \infty} \frac{\alpha(\omega)}{\omega}z$$

Under such condition, even if the rest of envelope propagates with a superluminal (negative group delay), the continuation of advanced phase of whole wave packet would give rise to no distortion of traversing wave packet.

Here we study the effect of varying the density of dielectric distribution on controlling group index via periodic geometrical transition. The suggested optical component provides a unique time evolving wave packet from a given initial state that accumulate advance phase for a broadband range of wavelength near the band edge. This class of mediums, modulated photonic crystals[13], with asymmetric nature of periodic structures exploits unbalanced density of states not only in first Brillouin zone but also through the entire lattice due to continuous broken symmetries. In summary, a smooth disordered hexagonal-like cluster at 2-dimension lattice is distributed through the entire suggested component in which the volume of air (cylindrical holes) changes only along($\Gamma M$). A Left handed material (LHM) with negative gradient of wave-vector ($k'(\omega) < 0$) forms in a periodic lattice to create NGD that we will discuss in this study.

**LHM versus RHM**

Light as a wave oscillates due to the fluctuation of two components ($\vec{E}\ and\ \vec{H}$) that ideally crosses each other at two trajectory planes which are perpendicular to each other. From complete wave point of view in an inhomogeneous matter with no spatial filtering, creation of a partial polarized light pulse can be affected by means of either horizontal or vertical components in case of asymmetric distributing in either one of fields through the media. Therefore, propagating such wave packets experience modification in both wavelength and magnitude (Figure 1-a). As it is clear from the figure, assuming one of the arrows (red or blue) may occurs in such transition. Thus, evolving ($\vec{E}\ and\ \vec{H}$) fields by considering conservation of energy are likely possible. In Figure 1-b, variation of only wavelength in an inhomogeneous composition such as a PhC creates two types of evolution from their initial states ($\omega_i$) that ends up with a particular angular frequency of whole envelope ($\omega_g$) which determines the group velocity of wave packet. In an special case, the coupled-mode theory (CMT) in a subwavelength scale based on the perturbation of refractive index through the lattice that explains possibility of producing a complete backward wave scattering due to optically linear parametric (OLP) interaction (Figure 1-c). So this gives rise to coalescence of of two eigenstates (Figure 1-d) near the band edge which possesses capability of diverse group delays. In photonics science, for an elaborate understanding of any types of dispersion in any region (linear or nonlinear) for given heterogeneous interfaces in (meta)-material, there should be a map to steer light in a dexterous manner. For instance computing band diagram is a necessary step to have a seminal record of EM wave interacting with any types of matter including homogeneous or inhomogeneous ones. Thus, instead of time-independent Schrodinger equation[14], using master equation which is so-called eigenvalue equation[15] produces eigenvalues of all possible states. By considering the linearity of Maxwell's equations for both fields in time, $\boldsymbol{E}(\boldsymbol{r},t) = \boldsymbol{E}(\boldsymbol{r})e^{i\omega t}\ and\ \boldsymbol{H}(\boldsymbol{r},t) = \boldsymbol{H}(\boldsymbol{r})e^{i\omega t}$, the solution via Fourier theory will not construct the possible solutions for its given components. Instead, combination of all propagating modes (Bloch envelop) as a solution for both fields can be retrieved through solving the given master-equation that is referred to each of the computed $\omega$

[16]. To have a clear image of how the propagation modes as a solutions for the Maxwell equation, one needs to eliminate the electric field **E** and apply the Bloch envelope for a broken symmetries in two fields and transverse guiding mode for only **H** field[16]:

$$\Theta\, H(r) = \left(\frac{\omega}{c}\right)^2 H(r)$$

Where $\Theta\, H(r) \equiv \nabla \times \left(\frac{1}{\varepsilon(r)} \nabla \times H(r)\right)$, exhibits functional form of dielectric to determine magnetic field. By considering a Hamiltonian system for a suggested system, since the Hermitian operator ($\Theta$) derives all eigenstates for magnetic fields, the determination of eigenstates for electric field would be the next step. Therefore, applying a translation of any symmetry such as a photonic crystal gives rise to the dispersion relation in which the splitting bands is expected.

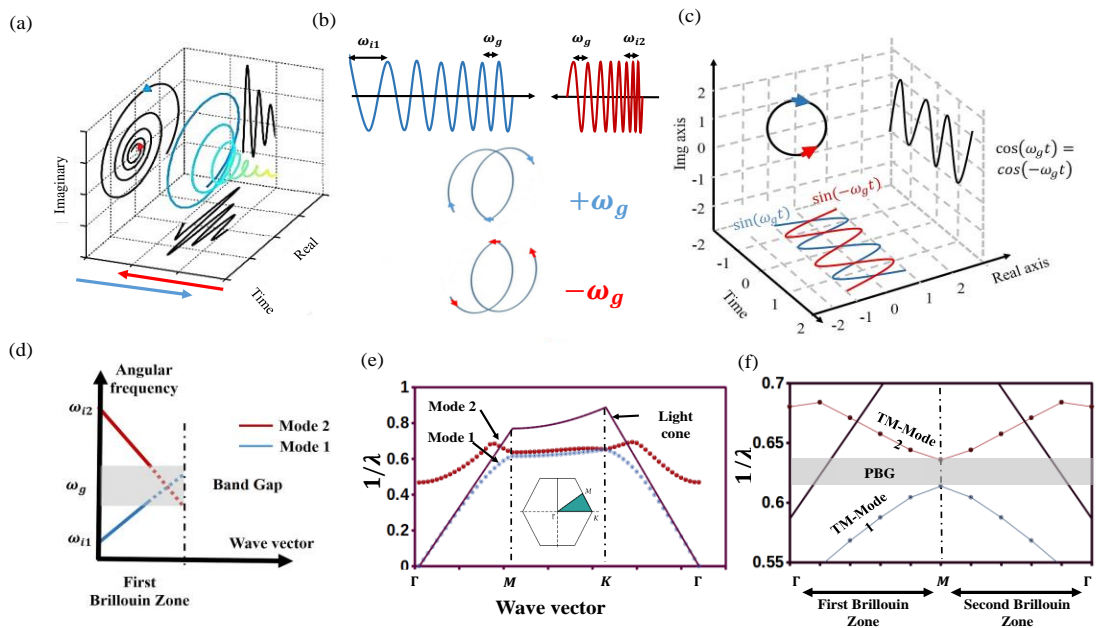

**Figure 1.** Interaction of waves (wave packet) and types of group delay; (a) time variation transition of wave packet through a real space form either in ascendant to descendant group velocities (b) 2-dimension wave transition with initial frequency ($\omega_i$) to the dominant group velocity ($\omega_g$) that can be either $+\omega_g$ or $-\omega_g$ which are different in direction of motions (c) trajectories of two possible envelopes which differs in one component and are the same in the other component (d) cancelation of two modes due to interacting two envelops with different group velocities in a normal dispersion fashion (e) The dimensionless quantity $\frac{\omega a}{2\pi c}$ (a=0.5 [μm] as a lattice constant) versus wave vector in a hexagonal lattice ($a = 0.5 \ [\mu m], Th_{Si} = 0.32 \ [\mu m], Th_{SiO2} = 1 \ [\mu m]$) (a) zoomed in band diagram in $\Gamma - M - \Gamma$ direction that shows symmetries in optical path with creation of photonic band gap (PBG).

Implementing Bloch's theorem via MPB[17] computational tools, we conduct a simulation with a high resolution FE method in which the grid elements for 3-dimension structure, hexagonal lattice, is applied. The computed result of photonic band structure for such lattice (cylindrical air) printed in a silicon film with the thickness of 0.32 [μm] on top of one micrometer silicon dioxide as a substrate is depicted Figure 1(e). As it is clear from figure, we calculate only TM-mode that is corresponded to the eigenmodes of the z-component for the applied electric field (**E$_z$**) in the proposed hexagonal lattice that is summarized at three directions (Γ, M and K) due to 2-dimensional symmetries at the first Brillouin zone. Furthermore, computed bands with the finite element method (FEM) shows power disintegration along photonic band gap (PBG) in the $\Gamma - M$ direction. However, the light flow is deviated to the other symmetrical direction due to joining computed bands at the K points. Thus, we expect a symmetrical band diagram in second Brillouin zone due to symmetric nature of the lattice in one of the directions (See the Figure 1-f). This diagram provides a seminal information for light steering in an engineered matter. For instance, by considering such photonic band structure, both LHM and RHM zone of a designed PhC can be recognized due to different gradient of wave vectors introduced around the PBG. In addition, this zone is subjected to geometric order and swings via any variation of designed parameters. For example, we can achieve swinging stop-band along the communication range of frequencies by means of either perturbation of lattice or changing one designing parameters such as lattice constant or radii. In Figure 2, the swinging stop band is depicted for both bands separately. Since gradient of wave vector $k(\omega)$ takes both negative and positive values in each portion of the PBG (positive values for first band and negative values for the second band) with respect to the direction of propagating in reciprocal lattice (e.g. in $\Gamma - M$), the type of dispersion follows these different values as well. According to the

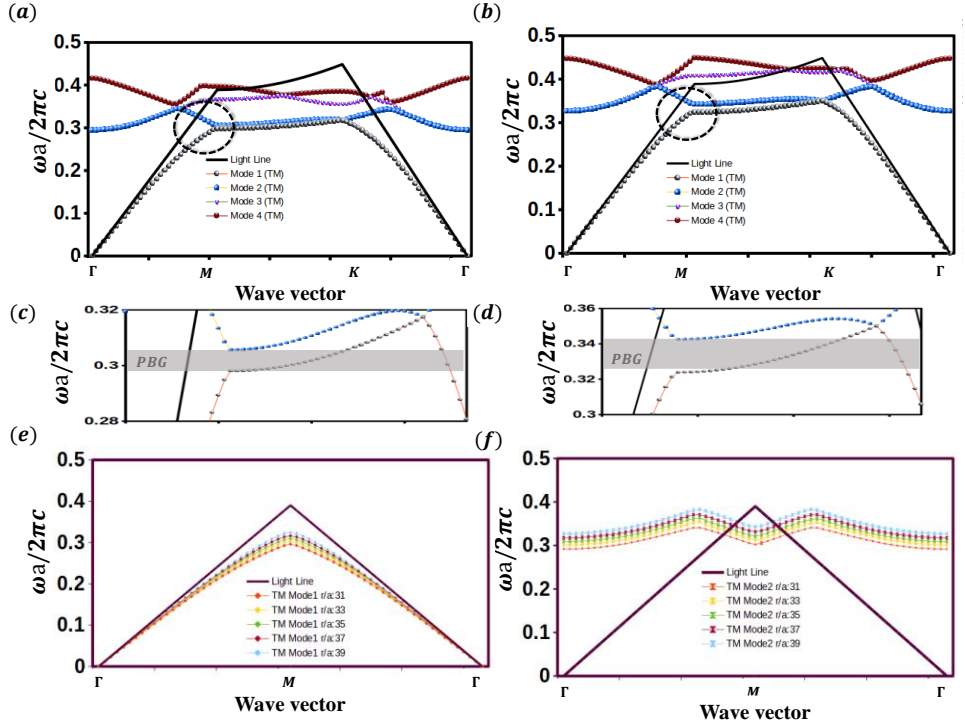

**Figure 2.** Swinging band diagram for a hexagonal lattice for the under study lattice (a) with the radii equals to 0.31 PBG occurs around 0.3 in $\Gamma - M$ direction (b) with the radii equals to 0.39 PBG occurs around 0.33 in $\Gamma - M$ direction (c), (d) zoomed in part of the photonic band gaps (e) swinging first band to the upper angular frequencies with increasing the radii (f) swinging second band to the higher angular frequencies with the increasing radii

However, we can examine this phenomenon from only transmission spectrum but band structure provides more specific information about the effect of changing any parameter on any of specific modes through the computed dispersion diagram. This also helps to calculate the gradient of wave vector (e.g. first and second order) with respect to the given frequency $\omega$ at both reciprocal and Cartesian spaces. Furthermore, we want to show that the effect of different values (either negative or positive) for gradient of wave vector in a PhC. In **Figure 3**, we provide a slice of unsteady simulation with FDTD method for 2-dimensional hexagonal lattice to specify differences in any types of chromatic dispersion. In other words, left-handed materials (LHM) and right-handed materials (RHM) are results of chromatic dispersion which both influence the angle of radiation due to order of phase velocity for propagating incident through the lattice. According to the Snell's law, the refracted incident from positive index material (PIM) to the negative index material (NIM) follows this equation $n_{PIM} * sin\theta_{PIM} = n_{NIM} * sin\theta_{NIM}$ and the backward bending of ray represents negative angle of refraction in the PhC which is depicted in **Figure 3**-b. Thus, satisfying the Snell's law with negative angle of refraction is possible only to multiply it with negative values of refraction indexes. As a result, the subscript of NIM and PIM is assigned to parameters in the equation for accurate analysis of chromatic dispersion in left-handed materials. In **Figure 3**-b, the nature of first order dispersion varies for the same light pulse profile but different radii since chromatic dispersion depends on gradient of wave vector as it follows the band diagram. So, any perturbation in geometry gives rise to swinging photonic band diagram and reconstruction of different values of group velocity $(k'(\omega) \equiv \partial k/\partial \omega = 1/v_g)$ is a possible method.

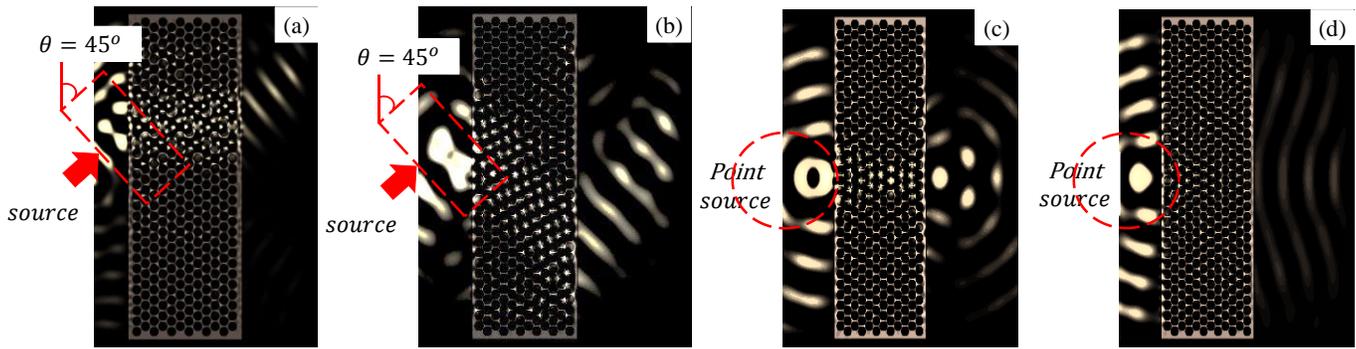

**Figure 3.** Light flow through the photonic crystal $(a = 0.5\ [\mu m], Th_{Si} = 0.32\ [\mu m], Th_{SiO2} = 1\ [\mu m])$ provides fundamental understanding about different types of light interacted with (a) RHM and tilted plane wave source $\theta = 45°$ (left) with respect to the length of the structure (b) LHM and tilted plane wave source $\theta = 45°$ (left) with respect to the length of the structure (c) LHM and point source (left) of light and its image in the right side of the structure (d) LHM and parallel plane wave incident (left)

Consequently, we conclude that changing the optical parameters in geometry of lattice such as radii or lattice constant gives rise to different types of first order dispersion. All in all, for a particular range of frequency the gradient of wave vector $(\frac{\partial k}{\partial \omega}, \frac{\partial^2 k}{\partial \omega^2})$ can be either negative or positive which influence the nature of dispersion. However, for creating a broadband negative refractive index to keep the entire envelop distortion-less one needs a dexterous approach to exploit geometric effect on photonic band diagram which we will discuss it in the following section.

**Modeling structure via asymmetric lattice**

Per our previous discussion, using MIT Photonic-Bands MPB as a well-known Eigenmode solver via Plane Wave Expansion (PWE) method to reach the accurate photonic band structure gives us a connection between given electromagnetic radiation and optical medium properties. It is worth to mention advantages of choosing a hexagonal lattice since it has capable of realizing band gap in both TM and TE modes simultaneously. In other words, proficiency of having two dimensional symmetry in such lattice, dielectric globs and connective veins establish a photonic band gap (PBG) to control light efficiently. In order to have deep understanding of originating a possible band gap in a suggested lattice one needs to be aware of some fundamental notions. First of all, distributed electric fields in a lattice with two components (air and dielectric) is in a way that the lowest order mode (first band) prefers to reside in higher index regions so it is called dielectric band. Unlike the dielectric band (first band), second order band prefers realization of electric field power in air (air band). Once distribution of these two modes become orthogonal to each other the creation of gap is possible due to the interference of two waves with the same periodicity for the same frequencies (See the discussion in Figure 1). Thus, some band frequencies cannot propagate at that particular direction. What if we break the periodicity via changing distribution of dielectric? For example, reducing or increasing the diameter of cylindrical holes (air) in 2-dimension lattice will breaks the periodic dielectric continuously. According to our former results, the band structure fluctuate with respect to such variation and gives rise to swinging in energy level with respect to the degrees of variation. We study the effect of such variation and proposed an approach to explain our novel modeling structure. In fact, the more symmetry is the more similar brag grating appear in computed directions. Otherwise, the cancelation of two bands will replace with the accumulation of phase. In other words, applying a linear variation of radii in each column (unit cell) at hexagonal lattice gives rise to dynamic variation of wave vector as it is flowing through the lattice. Let's assume a propagated incident in $\Gamma - M$ direction that the radii of unit cells in z direction varies ($\Delta r = D_2 - D_1$) smoothly to avoid any Fresenal reflection based on no optical impedance mismatch. Due to dependency of wave vector to changing radii, the gradient of wave vector with respect to the frequency ($\partial k(z)/\partial \omega$) will be either descendant (negative) or ascendant (positive). Thus, the type of chromatic dispersion can be varies as it is showed in Figure 4.

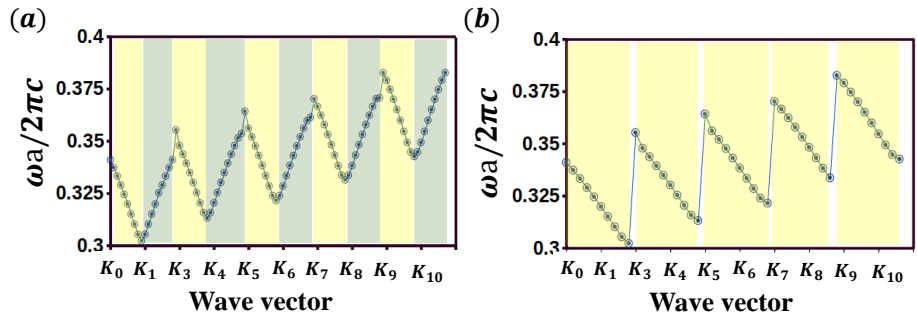

Figure 4. Dispersion diagram in asymmetric cluster (hexagonal lattice) through the light path of the modulated optical element (a) Discrete portion of band diagram in the first Brilouin zone of entire lattice (b) Gradient of wave vector in reciprocal space with continuous broken symmetries of lattice

We applied a Gaussian pulse to compute its transmission spectrum from both types including ascendant and descendant dielectric volume distributed in the lattice. Using Finite-Difference Time-Domain method in MIT electromagnetic equation Propagation (Meep)[18] to illustrate different optical properties in transmission spectrum for a particular range of frequency in both disordered lattice. The simulation is done in our suggested 3-dimension structure with 2-dimension photonic crystal that array of cylindrical holes (air) printed into silicon film with the thickness of 0.32 [μm] over one micrometer silicon dioxide as a substrate. First, we addressed asymmetry effect on variation of radii equal to 0.35 that experience smooth linear reduction of diameter to reach the both 0.33 (Figure 5-a) and starting radii from 0.35 to 0.37 (Figure 5-b) separately. Then, the computed evolved transmission spectrum in the both structures are compared with the one which is not experienced disordered effect (Figure 5-c). In Figure 5-d, the swinging stop band exhibits the difference in both advance and delay group velocity in frequency domain. However, one can conclude the difference of phase in two diagrams (blue and red) from the (Figure 5-c) in which the amplitude of transmission part in blue diagram and widening of stop band in the blue diagram of (Figure 5-d) both represents negative group delay of the envelope in frequency domain.

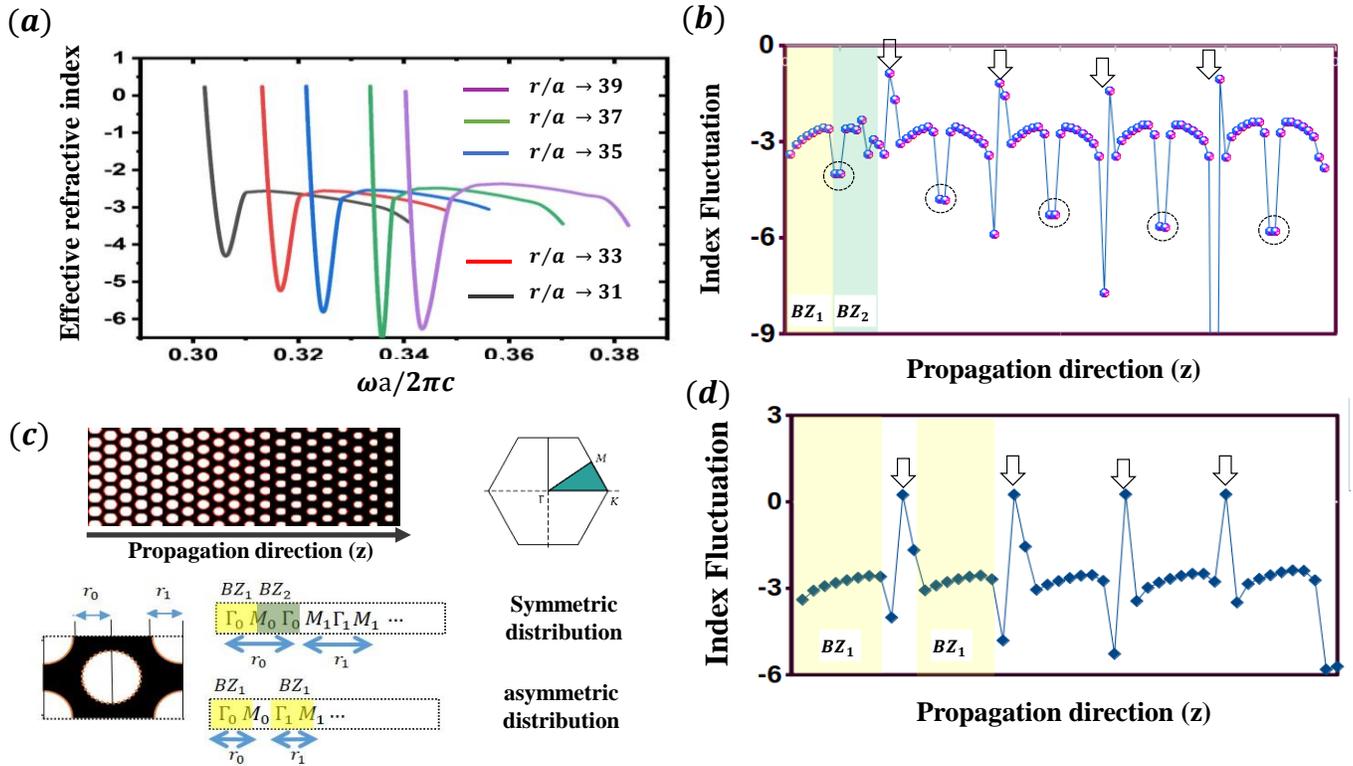

Figure 5. Computed results (a) effective refractive index for different radii values in an hexagonal lattice (b) extracted refractive indexes for symmetric hexagonal lattice in both first and second Brillouin zones for different radii values with ascending orders through the propagation direction (c) All details of proposed lattice in two types of distributions (symmetric and asymmetric) in which the asymmetric distribution without experiencing the second Brillouin zone traverses to

Our suggested metamaterial composed of multiple clustering hole meshes generates exotic band diagram due to broken symmetries. This platform can be used to control density of states near the band via continuous perturbed parameter (radii) and to widen the negative band in the communication wavelength. In Figure 5-a, refractive index with negative values spanning form $(a/\lambda) = 0.385$ which equals to $\cong 1.3\ \mu m$ to $(a/\lambda) = 0.31$ which equals to $\cong 1.6\ \mu m$. This width in telecom consists extended band (E), short band (S), conventional band (C) and long band (L). However, this broadband negative index region have functionality depending on the orientation ($\Gamma M$). Thus, this broadband functionality is based on connectivity rather than consequence of resonating electromagnetic wave in subwavelength structure. In Figure 5-c, the proposed lattice with continuous perturbing radii shows two types of distributions. One with symmetric hexagonal lattice which consists of two Brillouin zones and the other cluster which before entering to the symmetric second Brillouin zone light traverse to the perturbed stack. To understand the difference in types of distributing clusters, one can figures out their effective refractive index fluctuation through the light path ($\mathbf{\Gamma M}$) in Figure 5-(b, d). The perturbing effect in these two sections merely provides radii for 0.31 to 0.39 in 4 steps which is not what the real structure under study refers on it. In other words, the proposed structure experience perturbing radii with very smooth transitions of radii to have lengthy device for realistic applications.

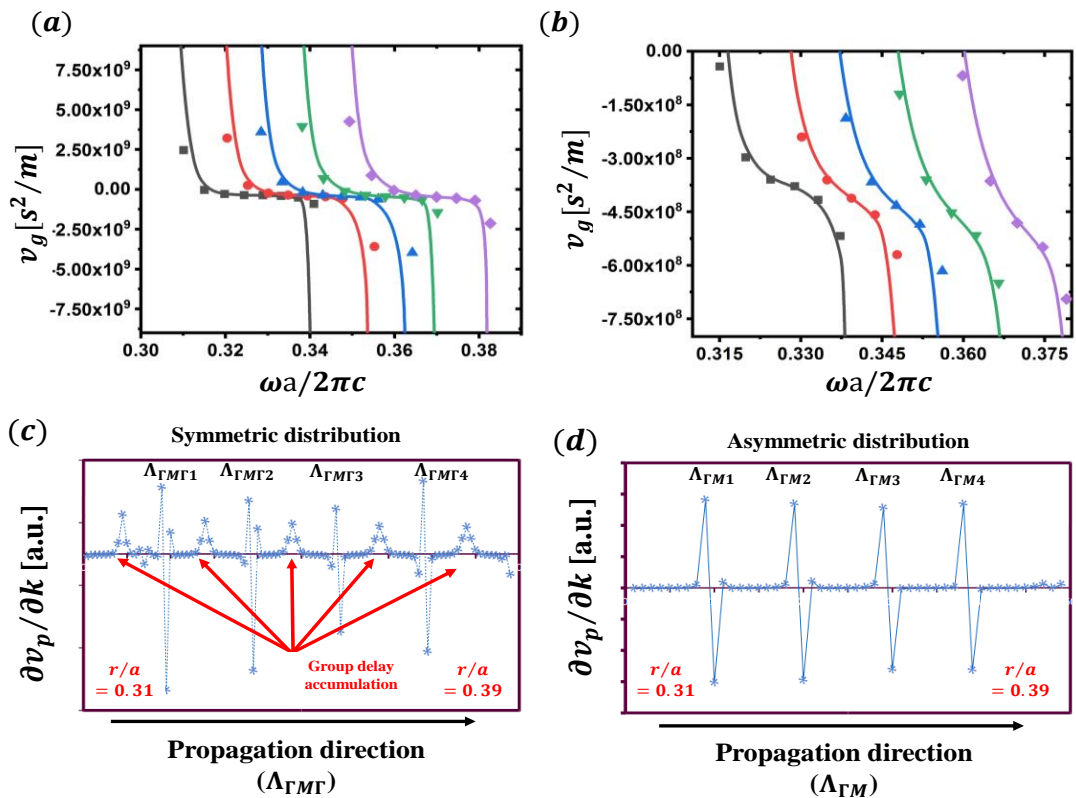

Figure 6. (a) Group velocity with swinging radii from 0.31 (black) to 0.39 (pink) (b) and its zoomed in region with negative group velocity (c) symmetric distribution which accumulate group delay in each transition from first Brillouin zone to the second one and weak compensating for transition of perturbing radii (b) asymmetric distribution with smooth group delay and compensating group delay in each transition of perturbing radii

By considering a broadband propagating, light pulses are distorted due to different phase velocities at each frequency. Thus, the complexity of signals at the end of light path is inevitable because of generating multiple intrinsic modes that each exhibits nonlinear group delay (GD). It is a fact that dispersion is a characteristic originating from the environment (media) and somehow depends on the properties of source as well. That is why characterizing the group delays extracting from each mode is considered as an indispensable which utilized as a parametric inversion to retrieve signal containing the information. Here we shows the difference of two suggested structures with specific group delays. In Figure 6, type of distributing clusters in a specific lattice, continuous perturbed radii, influence the group delay in any transition through the light path. According to the gradient of phase velocities of both symmetric and asymmetric distribution, the one which does not enter to the second Brillouin zone and experience perturbing effect promptly possesses a smooth group delay rather than the cluster with symmetric distribution. In our suggested structure, continuous perturbed lattice which is a proper platform for broadening the band with almost zero group delay, eliminating frequent positive values of group delay in each transition to the second Brillouin zone plays a destructive role. Furthermore, mismatching the phase of two clusters with different effective refractive index causes asymmetric values of GDs in each interval. Therefore, nonlinear accumulation of GDs at the end of light path gives rise to distorting light pulse in a complex manner. Unlike the symmetric distribution of cluster in PhCs, broken symmetries produce a proper platform which not only eliminate the accumulation of GDs in the middle of two Brillouin zones but also provide an almost symmetric values of GDs in each perturbed transitions which compensates each other.

**Conclusion**

This study introduced a new approach of tailoring photonic crystals to form a dynamic wave vector gradient through the continuous perturbed lattice. The exotic photonic band diagram gives rise to broadening of band in the negative refractive index region. This unique band diagram is due to connectivity of stacks rather than resonance in the subwavelength structure. This approach, smooth variation of dielectric distribution, extends photonic band edge toward upper or lower frequencies depending on types of disordered, either low to high (LH) or high to low (HL) density, and produces a broadband near zero group delay while the $k'(\omega, \vec{r}) < 0$.

Furthermore, we compared the two types of perturbing radii through the proposed lattice. We realized the both symmetric and asymmetric hexagonal clusters via modifying unbalanced distribution of dielectric through the entire component. In addition, eliminating accumulation of group delay occurs to a broadband via applying prompt perturbing phenomenon before traversing light pulse to the symmetric second Brillouin zone. The suggested optical component and the reported approach provides a promising technique to obtain a unique platform for producing near zero group delay (NZGD) with covering a broadband range of frequencies adjacent to the photonic band diagram with keeping the power flow away from dissipation through the transmission line.